\documentclass[preprint,journal]{vgtc}       

\ifpdf
  \pdfoutput=1\relax                   
  \usepackage{graphicx}                
  \DeclareGraphicsExtensions{.pdf,.png,.jpg,.jpeg} 
\else
  \usepackage{graphicx}                
  \DeclareGraphicsExtensions{.eps}     
\fi%

\graphicspath{{figures/}{pictures/}{images/}{./}} 

\usepackage{microtype}                 
\PassOptionsToPackage{warn}{textcomp}  
\usepackage{textcomp}                  
\usepackage{mathptmx}                  
\usepackage{times}                     
\usepackage{cite}                      
\usepackage{tabu}                      
\usepackage{multirow}
\usepackage{booktabs}                  
\usepackage{fontawesome}
\usepackage[dvipsnames,table]{xcolor}
\usepackage{fancyvrb}
\usepackage{float}
\usepackage{dblfloatfix} 
\usepackage{hyperref}
\usepackage[normalem]{ulem}

\definecolor{removed}{RGB}{222,222,222}
\definecolor{edit}{RGB}{255,127,14}

\newcommand{\revision}[1]{\textcolor{edit}{#1}}
\renewcommand{\revision}[1]{#1}
\newcommand{\cut}[1]{}

\definecolor{tableau1}{RGB}{31,119,180}
\definecolor{tableau2}{RGB}{255,127,14}
\definecolor{tableau3}{RGB}{127,127,127}
\definecolor{tblgrey}{RGB}{222,222,222}

\newcommand{\first}[1]{\textcolor{tableau1}{#1}}
\newcommand{\second}[1]{\textcolor{tableau2}{#1}}
\newcommand{\grfl}[1]{\textcolor{tableau3}{#1}}

\newcommand{\ie}{i.e.,\ }
\newcommand{\etal}{et al.}
\newcommand{\eg}{e.g.,\ }

\newcommand{\bstart}[1]{\vspace{1mm} \noindent{\textbf{#1.}}}

\usepackage{makecell}

\preprinttext{Authors' version of the paper accepted to IEEE VIS 2021; to appear in IEEE Transactions on Visualization \& Computer Graphics.}

\onlineid{1065}

\vgtccategory{Research}
\vgtcpapertype{Theoretical \& Empirical}

\title{\revision{From Jam Session to Recital}: Synchronous Communication \\ and Collaboration Around Data in Organizations}

\author{Matthew Brehmer and Robert Kosara}
\authorfooter{
\item
 Matthew Brehmer and Robert Kosara are with Tableau Research. E-mail: \{mbrehmer,rkosara\}@tableau.com.
}

\shortauthortitle{Brehmer \& Kosara: From Jam Session to Recital}

\abstract{Prior research on \revision{communicating with visualization} has focused on public \revision{presentation} and asynchronous individual consumption, such as in the domain of journalism.
The visualization research community knows comparatively little about synchronous and multimodal communication around data within organizations, from team meetings to executive briefings. 
We conducted two qualitative interview studies with individuals who prepare and deliver presentations about data to audiences in organizations.
In contrast to prior work, we did not limit our interviews to those who self-identify as data analysts or data scientists.
Both studies 
\revision{examined aspects of} speaking about data with visual aids such as charts, dashboards, and tables. 
One study was a retrospective examination of current practices and difficulties, \revision{from which we identified three scenarios involving presentations of data.
We describe these scenarios using an analogy to musical performance}: small collaborative team meetings are akin to \textsl{jam session}, \revision{while more structured presentations can range from \textsl{semi-improvisational performances} among peers to formal \textsl{recitals} given to executives or customers}. 
\revision{In our second study, we grounded the discussion around three design probes}, each examining a different aspect of presenting data: the progressive reveal of visualization to direct attention and advance a narrative, visualization presentation controls that are hidden from the audience's view, and the coordination of a presenter's video with interactive visualization.
\revision{Our distillation of interviewees' responses surfaced twelve themes, from ways of authoring presentations to creating accessible and engaging audience experiences.}
} 


\CCScatlist{ 
 \CCScat{K.6.1}{Management of Computing and Information Systems}%
{Project and People Management}{Life Cycle};
 \CCScat{K.7.m}{The Computing Profession}{Miscellaneous}{Ethics}
}

\vgtcinsertpkg

\begin{document}


\firstsection{Introduction}

\maketitle

\label{sec:intro}

Meetings and presentations are a routine part of
\revision{life for people working in organizations},
and \revision{many of these events} involve visual displays of data.
Yet the visualization research community knows little about how the data is presented, the intent behind these presentations, 
or how they
differ from the better-understood scenarios of data analysis and asynchronous document-centric communication around data. 

The act of presenting charts, tables, and dashboards as visual aids to support spoken narratives and discussions are central to the studies presented here. 
We refer to these as \textit{performative} presentations because they require a presenter to narrate and step through the content; we look beyond the preparation of slides or other materials for asynchronous consumption. 
These presentations can be in person or remote, live or recorded. Ultimately, across all cases, at the center is a person with an intent to communicate data, results, or insights to an audience.

Rather than being an end in and of itself, a presentation is often a starting point for collaboration. 
This includes collaborative sensemaking around new data, the solicitation of explanations that add context to data from a team, or deeper conversations and debates driven by the audience's questions. 
The tools used to present data are widely used and known, but they are poorly suited for synchronous performative presentations. 
Visualization tools have only rudimentary support for live presentation, and visualization support in presentation tools (or \textit{`slideware'}) is similarly limited. 
Currently, screenshots are the common conduit between these tools, bringing with them a loss of interactivity and connection to the underlying data~\cite{elias2012annotating,sarikaya2018we}.
There is little prior research in the visualization community addressing these scenarios within an organizational setting; our research sheds light on this under-addressed area and represents a call to action for researchers and tool builders, particularly as more organizations undergo a process of digital transformation~\cite{turco2016conversational} and adopt practices of communication and collaborative decision-making grounded in data.

We conducted two interview studies with 18 individuals who prepare and deliver presentations about data to audiences within organizations.
In contrast to previous interview studies that examined collaboration around data within organizations, our interviewees were not limited to data analysts or data scientists.
One study included thirteen retrospective interviews and focused on existing approaches, pain points, and desires with respect to presenting data, in which we asked participants to show us presentation materials that exemplified their current processes.
Our second study included ten interviews, in which we used three design probes to elicit discussion around three aspects of presenting data to an audience: the progressive reveal of data, the use of a secondary display to modify how the audience sees the data, and the coordination of video with interactive visualization. 

This paper is a step toward understanding conversations around data that are common within organizations today.
Our primary contribution is \revision{our analysis of interviewees' responses} from both studies, providing the visualization community with an initial understanding of how individuals collaborate synchronously around data within organizations via the act of performative presentation. 
\revision{Our analysis led us to identify three scenarios involving the presentation of data that we describe using an} analogy to musical performance: if a small collaborative team meeting is akin to a \textit{jam session}, \revision{more structured presentations span the gamut from \textsl{semi-improvisational performances} among colleagues to formal \textsl{recitals} addressed at executives or customers}. 

We also contribute the three design probes used in the second study, as they demonstrate presentation functionality that is not present in existing visualization or presentation tools. 
Finally, we suggest a set of research and development opportunities that, if realized, could support the presentation of data to live audiences.

\section{Background and Related Work}
\label{sec:rw}

We build upon prior work on presenting and collaborating around data. 

\bstart{Performative presentations of data \textit{`in the wild'}}
Arguably the most notable performative presentations of data are Hans Rosling's 2006 and 2007 TED conference talks~\cite{Rosling2006,Rosling2007}.
Recordings of these performances have motivated a thread of visualization research projects beginning with a 2008 study by Robertson~\etal~\cite{Robertson2008}, which compared the relative efficacy of animation and small multiples.
Beyond this comparison, this study also found that an accompanying spoken narration has a positive effect on participants' comprehension of trends.
However, the study was unable to replicate other aspects of Rosling's performance: the tone and prosody of his monologue, his exaggerated gesticulations, and his stage presence before a large live audience.
It is important to acknowledge that few people can match Rosling's talent for presentation, and that well-rehearsed TED conference presentations are unrepresentative of most live presentations, whether or not they involve the presentation of data.
Though many recordings and presentation materials (\ie slides) from live presentations about data can be found online, from academic lectures~\cite{gomez2012different} to conference presentations from the \textit{quantified self} community~\cite{choe2015characterizing}, these are intended for \textit{public audiences}.

Similarly, performative presentations about data involving the use of visualization are increasingly appearing in televised newscasts. 
Despite this prevalence, we echo Drucker~\etal~\cite{drucker2018}'s observation that the presentation of visualization on television is \textit{``almost completely overlooked in visualization research,''} aside from being a common source of deceptive visualization examples~\cite{pandey2015deceptive}.

While an analysis of \revision{recorded conference presentations and televized performances involving data} would be illuminating,
\revision{we would still lack} insight about the dynamics of presentations delivered to live audiences where interruptions and discussions routinely arise, such as in the organizational contexts where we focus our current research.
\revision{Prior to our interview studies, we hypothesized that not all presentations involving data within organizations are formal performances, though we were not yet able to articulate the characteristics of other scenarios.}
\revision{Looking to prior work, we noted that Kandogan~\etal~\cite{kandogan2014data}'s interview study of data analysts hinted at the use of tables and simple graphical charts in enterprise presentations}, at least among mid-level and junior analysts.
Erete~\etal~\cite{erete2016storytelling} similarly alluded to how people present data within non-profit organizations, however they do not discuss the \revision{specific context} of how these presentations are delivered.
More recent analyses~\cite{crisan2020passing,zhang2020data} of data science roles and activities within organizations described knowledge dissemination and communication processes, encompassing report generation and presentation, however there is a dearth of writing describing how the latter set of activities are performed, effectively or not, within these settings. 
The roles of our interviewees were more heterogeneous, going beyond data analysts and data scientists, with titles that included sales executive, marketing manager, and communications specialist.

\bstart{Narrative visualization}
Performative aspects have also permeated into previously non-performative forms of narrative visualization~\cite{segel2010narrative,riche2018data} that are consumed asynchronously, namely online journalism.
Despite early enthusiasm for interactive visualization experiences in which sequential reveals and transitions were triggered by a viewer's interaction, researchers and practitioners alike discovered that most viewers do not engage with such functionality~\cite{boy2015storytelling}, prompting news graphics teams to incorporate less interactivity in their work ~\cite{Aisch2016,Tse2016}.
However, a sequential reveal of information is nevertheless invaluable to narrative visualization~\cite{hullman2011visualization,hullman2013deeper,hullman2017finding} and data journalism~\cite{Fairfield2015}, whereby these reveals can manifest in both time and space.
In response, one approach has been to better align reveals and transitions in visualization content with a viewers' reading position in an adjacent text article~\cite{mckenna2017visual}. Another approach is performative, treating narrative visualization as a form of documentary filmmaking~\cite{amini2015understanding,tang2020narrative}, incorporating rich visual vocabularies, musical scores, and voice-over narration~\cite{Halloran2016}.
Despite authoring tools such as Idyll~\cite{conlen2018idyll}, which allows for interleaving the reveal of visualization and text, or Flourish~\cite{FlourishStudio}, which includes functionality to add voice-over narration to a visualization reveal sequence~\cite{FlourishTalkies}, the process of authoring these presentations is time-intensive and typically involves the use of multiple tools.
Moreover, these narrative visualization tools cater primarily to asynchronous consumption, and the discussions that takes place among readers in article comment sections rarely address the data being presented~\cite{mcinnis2020rare}.
In contrast, our interviews and design probes speak to the unique needs of revealing information in presentations delivered to live audiences. 

\bstart{The ubiquity of slideware}
While the research community is well aware of the ubiquity of slide presentation software such as PowerPoint, Keynote, and Google Slides in academic lectures~\cite{gomez2012different}, this software is similarly ubiquitous in organizational settings.
Critiques of PowerPoint and similar \textit{slideware} have prompted exploration into alternative presentation tools that aspire to add flexibility and non-linearity to the stages of presentation preparation, rehearsal, and delivery~\cite{edge2013hyperslides,edge2016slidespace}, or to tools that blur the distinction between presentation and collaborative discussion~\cite{chattopadhyay2018beyond,muaruacsoiu2016clarifying,tian2021system}. 
Despite these advances, requirements specific to presenting data remain salient in this nascent class of post-PowerPoint tools.
Beyond the issues of low information density and a small expressive gamut for presentation-oriented representations of data~\cite{kosara2016presentation,tufte2003cognitive}, a third issue that we address pertains to the static nature of media assets that presentation tools import, resulting in an inability to flexibly control dynamic or interactive content at different levels of granularity during the delivery of a presentation.
We revisit these requirements in the context of the presentation scenarios (\autoref{sec:scenarios}) and as aspects of our design probes (\autoref{sec:probes}).

\bstart{Collaborative visualization}
\revision{The design space for collaboration around visualization~\cite{isenberg2011collaborative,viegas2006communication} encompasses the dimensions of time (synchronous vs. asynchronous) and space (co-located vs. remote).
In this paper, our focus is primarily on synchronous collaboration and our characterization of presentation in relation to it}.

\revision{Prior research on co-located synchronous collaboration around data includes studies of team-based sensemaking and analysis tasks, such as where team members interact with a common visualization tool either individually via a personal terminal~\cite{mahyar2014supporting}, as pairs with a shared large-display terminal~\cite{vogt2011co}, or as a group around an interactive tabletop display ~\cite{isenberg2011co}.
There also exists a body of research investigating the potential of large and interactive upright displays~\cite{knudsen2019pade,langner2018multiple} for supporting synchronous collaboration around data, including those that afford both touch- and pen-based interaction~\cite{lee2013sketchstory,zgraggen2014panoramicdata}, which could be used for presenting data in an engaging way to a co-located audience.
}

\revision{Recent work by Neogy~\etal~\cite{neogy2020representing} and Schwab~\etal's VisConnect system~\cite{schwab2020visconnect} both addressed remote synchronous collaboration, with the latter suggesting applications and roles associated with telemedicine (doctor and patient) and online education (teacher and students).
We are also interested in role differentiation, such as where a designated presenter or moderator operates a software tool while others ask questions, offer explanations, or make predictions based on their experience or expertise.}
This scenario recalls Arias-Hernandez~\etal's notion of paired analytics~\cite{arias2011pair}, in which an analyst with expertise in visual analysis software works alongside a subject matter expert; we extrapolate this scenario to small groups and assume that the visual analyst has already conducted some exploratory data analysis prior to meeting with the subject matter experts.

Our findings also address the collaborative preparation of presentations about data, recalling Loorak~\etal's work on asynchronous collaborative visualization authoring~\cite{loorak2018changecatcher}, which assumed collaborators working with a shared software environment.
In reality, we find that this collaborative authoring is more complicated, involving hand-offs between tools \textit{and} people, even across organizational boundaries.

\newcommand{\X}{\grfl{\faSquare}}
\begin{table*}[!tp]
  \caption{\revision{Aspects of presenting data across interviewees (* = we conducted a joint retrospective interview for colleagues P05 and P06)}. \revision{Multiple frequencies may correspond with different scenarios and/or presentations to different audiences}. Colors indicate \revision{participation across the two studies}: \first{\faSquare} = retrospective, \second{\faSquare} = design probe. \revision{Scenario abbreviations: \textit{\textbf{J}am Session}, \textit{\textbf{S}emi-Improvised Performance}, \textit{\textbf{R}ecital}.}
  }
  \label{tab:participants}
  \small%
	\centering%
  \begin{tabu}{%
  ll|%
  *{3}c|%
  *{3}c|%
  *{3}c|%
  *{3}c|%
  *{3}c|%
  *{3}c%
  }
  & & P01 & P02 & P03 & P04 & P05$^*$ & P06$^*$ & P07 & P08 & P09 & P10 & P11 & P12 & P13 & P14 & P15 & P16 & P17 & P18  \\
  \midrule
\parbox[t]{1mm}{\multirow{2}{*}{\rotatebox[origin=c]{90}{Study}}}  & \revision{First interview} & \second{\faSquare} & \first{\faSquare} & \first{\faSquare} & \first{\faSquare} & \second{\faSquare} & \first{\faSquare} & \first{\faSquare} & \first{\faSquare} & \first{\faSquare} & \first{\faSquare} & \first{\faSquare} & \first{\faSquare} & \first{\faSquare} & \first{\faSquare} &  \second{\faSquare} &  \second{\faSquare} & \second{\faSquare} & \second{\faSquare}  \\
  
 & \revision{Second interview} & \first{\faSquare} & \second{\faSquare} & & & \first{\faSquare} & & & & & & \second{\faSquare} & & \second{\faSquare} & \second{\faSquare} & & & &  \\
\midrule
 \parbox[t]{1mm}{\multirow{2}{*}{\rotatebox[origin=c]{90}{Who}}} 
 & Deliver personally & \X & \X & \X & \X & & & \X & \X & \X & \X & \X & \X & \X & \X & \X & &  & \X \\
 & Prepare for others & & & \X & & \X & \X & & & \X & & & & & & & \X & \X &  \\
\midrule
\parbox[t]{1mm}{\multirow{3}{*}{\rotatebox[origin=c]{90}{\revision{When}}}} 
& Daily to weekly & & \X & \X & \X & & & & & & \X & \X & \X & \X & & & & & \\
& Monthly to quarterly & \X & & \X & & \X & \X & \X & \X & \X & & & & & & \X & & \X & \X \\
& Yearly / irregularly & \X & \X & & \X & & & \X & & \X & \X & & & & \X & & \X & \X & \X \\
\midrule
\parbox[t]{1mm}{\multirow{5}{*}{\rotatebox[origin=c]{90}{Audience}}} 
& Immediate team & & & & \X & & & & & \X & & & \X & \X & & & & & \\
& Immediate management & \X & \X & \X & \X & & & & & & & \X & \X & & & & & & \\
& Cross-functional teams & & & & & \X & \X & & & & & & & \X & \X & & \X & \X & \X \\
& Executives, owners  & & & & & & & \X & \X & \X & \X & & & \X & & & \X & \X & \X \\
& Customers, clients & & \X & \X & & & & \X & \X & & \X & & & & \X & \X & & & \\
\midrule
& \revision{Scenarios} & S & S+R & J+R & J+S & R & R & R & S+R & J+R & R & J & S & S & R & S & J & R & R \\
\midrule
& Industry & \parbox[t]{1mm}{\rotatebox[origin=r]{90}{Software}} & \parbox[t]{1mm}{\rotatebox[origin=r]{90}{Law}} & \parbox[t]{1mm}{\rotatebox[origin=r]{90}{Trade Assoc.}} & \parbox[t]{1mm}{\rotatebox[origin=r]{90}{Consulting}} & \parbox[t]{1mm}{\rotatebox[origin=r]{90}{Software}} & \parbox[t]{1mm}{\rotatebox[origin=r]{90}{Software}} & \parbox[t]{1mm}{\rotatebox[origin=r]{90}{Education}} & \parbox[t]{1mm}{\rotatebox[origin=r]{90}{Education}} & \parbox[t]{1mm}{\rotatebox[origin=r]{90}{Retail}} & \parbox[t]{1mm}{\rotatebox[origin=r]{90}{Health}} & \parbox[t]{1mm}{\rotatebox[origin=r]{90}{Research}} & \parbox[t]{1mm}{\rotatebox[origin=r]{90}{Finance}} & \parbox[t]{1mm}{\rotatebox[origin=r]{90}{Consulting}} & \parbox[t]{1mm}{\rotatebox[origin=r]{90}{Finance}} & \parbox[t]{1mm}{\rotatebox[origin=r]{90}{Consulting}} & \parbox[t]{1mm}{\rotatebox[origin=r]{90}{Finance}} & \parbox[t]{1mm}{\rotatebox[origin=r]{90}{Software}} & \parbox[t]{1mm}{\rotatebox[origin=r]{90}{Software}} \\
  \end{tabu}%
\vspace{-3mm}
\end{table*}

\section{Interview Studies}
\label{sec:methodology}

\revision{The design probe interview study began before the retrospective interview study. 
After the first six design probe interviews, we noted the diversity of presentation scenarios in interviewees' responses to introductory questions. 
We therefore decided to probe deeper into the spectrum of presentation scenarios with a retrospective interview study. 
From this point onward, the two studies proceeded in parallel. 
}

\bstart{Interviewees} 
We conducted 23 interviews with 18 individuals 
with six participating in both studies (\autoref{tab:participants}). 
We recruited them from three sources: seven were referrals from our professional network, 
seven participated in a customer research event at \revision{Tableau Conference 2020},
and four were found on User Interviews~\cite{userinterviews}, an online research recruitment service.
\revision{Those recruited via the latter two sources completed a survey that asked them to select from a list of statements about their use of visualization tools. 
We accepted participants who selected ``\textit{I present visualizations or dashboards in meetings},''
``\textit{I create visualizations or dashboards and share them with others to use in presentations},'' or
``\textit{I record video presentations featuring visualizations or dashboards for others to watch later}.''}
\revision{Overall, our recruitment approach reflects the opportunistic nature of industry-situated research involving the study of working professionals, which includes referrals, snowball recruiting, and conducting interviews at events where these professionals gather.}

\revision{Collectively}, our interviewees represented several sectors of the economy, including software, manufacturing, education, private equity, retail consulting, law, and healthcare.
We also attained representation across the dimensions of gender (7 female, 11 male), geography (North America, Europe, and South America), and \revision{years of professional experience} (\revision{2--27 years, average 9}), from junior roles to a regional vice president of sales at a multinational software company.
\revision{After our first few interviews, we noted that those who prepare presentations are not always those who deliver them, and that the latter may not work alone, recalling Lee~\etal's classification of \textit{analyst}, \textit{scripter}, \textit{editor}, and \textit{presenter} roles~\cite{lee2015more} when making and telling data stories. 
Accordingly, we were curious to study the dynamics of those who work together in these capacities: P17 prepares presentation materials for P18, while P05 and P06 routinely work together to prepare materials that are presented by an executive vice president of sales; we therefore conducted a joint retrospective interview with the latter dyad.}

\bstart{Format}
\revision{We conducted and recorded all of our interviews via video conference}. \revision{Each session was led by one interviewer with an observer present to take notes and ask additional questions. Some of the sessions were joined by an additional observer from a product or user research team.}
The interviews lasted one hour with the exception of the retrospective interviews with P11--P14; these were only half an hour in length as they took place during a time-bounded conference program. 
However, we resumed our conversations with P11, P13, and P14 in design probe interviews after the conference.

The format for the retrospective interview was structured around a set of open-ended questions about the presentations that the interviewee had prepared or delivered. 
As they responded, we encouraged them to share their displays and show us their presentation materials so that we could ask grounded questions about how they presented data. We discuss the results from these interviews in Sections~\ref{sec:scenarios} and~\ref{sec:characteristics}.

The format for the design probe interviews began with roughly fifteen minutes of open-ended questions about current practices, being a compressed version of the retrospective interview format.
We then spent roughly forty minutes demonstrating and collecting \revision{open-ended} feedback on our three design probes. 
We describe the probes and the results from this part of the study in \autoref{sec:probes}. 

We include \revision{our interview scripts} for both studies along with the design probe materials as supplemental material.

\bstart{\revision{Data Analysis}} \revision{In addition to observer notes, our video conferencing software automatically generated transcripts. Both authors performed thematic analyses on notes and transcript quotes using spreadsheet and affinity diagramming tools. 
The themes of the retrospective study are captured the subsections of \autoref{sec:characteristics}, while the themes of the  design probe study are reflected by the bold paragraph titles captured within each subsection of \autoref{sec:probes}.
Finally, the three scenarios identified in the next section are based on a consolidated analysis of responses from both interview studies.
}

\section{Presenting Data: Three Scenarios}
\label{sec:scenarios}
\revision{Our analysis of interviewees' responses led us to identify three scenarios involving the presentation of data within organizations.}
Although we color our descriptions of these scenarios with quotes from individuals, many interviewees described more than one context in which they prepare or deliver presentations about data (see \autoref{tab:participants}). 
We therefore present \textit{scenarios} rather than \textit{personas}, in contrast to prior enterprise-centric interview studies~\cite{kandel2012enterprise,zhang2020data}. 
The extent and frequency of preparing and delivering presentations about data vary considerably across our interviewees, and their job titles and responsibilities beyond these activities were similarly varied.

\revision{Upon identifying the scenarios, we sought a succinct way to refer to them, leading us to our musical performance analogy:}

\bstart{\textit{Jam sessions}} These are casual and often improvisational performances among small groups of musicians. 
They are opportunities \revision{to iterate on new ideas and} rehearse for more \revision{structured} performances.  
A participant can come prepared with a set of melodies, rhythms, or key signatures, or they can respond to the suggestions of other players.
Analogously, in small collaborative team meetings around data (\autoref{sec:scenarios:small}), a single participant can prepare materials to share and review, which could include slides, documents, or charts generated \revision{from the use of} visualization tools. 
Others then might chime in with their own thoughts, views, questions, and additional data.

\bstart{\revision{\textit{Semi-improvisational performances}}} 
\revision{Improvisation in jazz concerts is based on prepared melodies, motifs, and modes. 
Unlike a jam session, some parts have been written before an improvisational performance, with specific sections carved out for improvisation that builds upon prepared material. 
The performers can respond to the mood of the evening or the audience, but they nevertheless have a prepared repertoire.}
This scenario corresponds to intra-organizational briefings that are are less frequent and more rehearsed than jam sessions (\autoref{sec:scenarios:medium}), though they often elicit some audience interaction.

\bstart{\textit{\revision{Recitals}}} Being the least collaborative scenario, materials and scripts for recitals are often prepared collaboratively. 
These formal presentations are prepared by multiple people over potentially long periods of time \revision{and} are extensively rehearsed (\autoref{sec:scenarios:formal}).

\revision{In summary, the key differentiating aspects of these scenarios are that of audience interactivity, the presence or absence of narrative linearity, and formality.}
\revision{These three scenarios can be alternatively characterized using the lens of Isenberg~\etal's~\cite{isenberg2011collaborative} levels of engagement for collaborative visualization. 
Across all scenarios, the audience is \textit{viewing} the presented material. 
As we move from recitals to semi-improvisational performances, audiences are \textit{interacting} with the presenter to a greater extent, whereas in some jam sessions, they may be interacting directly with shared presentation materials.
Finally, jam sessions can also be an opportunity for collaborative \textit{creating}, as we will encounter with the example of P03 profiled below in \autoref{sec:scenarios:small}.}

We further identified recurring themes spanning these scenarios, particularly around presentation frequency, tool use, and the visualization design choices presenters make, which we discuss in \autoref{sec:characteristics}. 

Three of our interviewees also reported preparing communicative yet non-performative documents presenting data with visualization and tables, including white papers, information graphics, and standalone slide decks to be consumed without a presenter. 
While broadly related to communicative visualization, we opt not to focus on this related scenario in this paper.

\subsection{Small Collaborative Team Meetings as \textsl{Jam Sessions}}
\label{sec:scenarios:small}

Meetings between two or more immediate colleagues are a common presentation scenario. 
One participant typically acts as a primary presenter or moderator, with other participants occasionally preparing and sharing additional content during the meeting. 
The specific goal of these meetings can vary, but P18 (a regional sales executive for a software company) succinctly captures an optimal experience for the presenter in these contexts: \textit{``if you're presenting in front of a group of peers} [\ldots] \textit{, you're trying to get to the point where people are engaged enough in the data to ask questions.''} 

We first consider the case of a subordinate meeting with their supervisor.
P03, an analyst for a manufacturing association, routinely prepares presentation materials about economic indicators using a business intelligence tool for his manager, which they review together in a room equipped with a large display.
These materials are the basis for more formal presentations that the manager, or sometimes the head of their organization, delivers to policymakers and members of the association at large, approximately every two or three months.
In the context of these one-on-one meetings, however, the presentation delivery is seldom linear or unidirectional, with P03's manager often requesting control of the mouse.
Instead, these meetings could be seen as conversations, where the goal is to identify presentation materials that they can later use to \textit{``prove} [their] \textit{point with charts.''} 
Often accompanying the charts P03 shows are his text descriptions of model assumptions.
However, there is little concern for information density or aesthetics in the presentation materials used in these conversations, as the preparation of slides for the more formal presentations delivered by P03's manager are outsourced to an independent design team.

Another small team dynamic is a division of labor and expertise, where one member conducts an initial analysis of a problem or a dataset and then serves as the team's guide.
This describes P04, an analyst at a consulting firm who examines a client firm's retail data with the goal of building retail performance forecast models. 
She conducts her exploratory data analysis using a business intelligence tool, which she also uses as the tool to present her findings to her team.
Such team meetings are highly interactive and ad-hoc, and in P04's case, occurring several times a week, and sometimes multiple times a day.
The goal of these meetings is to show her colleagues \textit{``the hallway, metaphorically speaking,}, [\ldots] \textit{showing them the doors that are available that I found in the data, and then they would open} [them]\textit{.''} 
These meetings are part of the initial data intake from a client; they help clarify questions and allow P04 to make decisions about which modeling steps to consider. 

Collaborative small team meetings need not be with a consistent team.
As an example, we consider P02, a lawyer in a large law firm with a background in computer science. 
He routinely meets with colleagues in his firm as well as with clients, discussing analyses he has performed in R, Python, and Tableau. 
He often prepares charts of time series data for these discussions, citing two recent examples: an employment law case pertaining to the timing of breaks for call center workers and a harassment case in which he performed an analysis of when those involved were co-located in the workplace based on employee badge reader logs. 
He often has his analysis software at his disposal during these discussions, and he mentioned adjusting parameters, re-running analysis scripts, and updating data sources in response to the discussion, enabling him to show updated charts to his colleagues or clients.

\subsection{Briefings as \revision{\textsl{Semi-Improvised Performances}}}
\label{sec:scenarios:medium}

Further along the spectrum of formality are presentations to groups beyond one's immediate colleagues, or in more hierarchical organizations, presentations for audiences further up one's management chain.
Unlike meetings among one's immediate team, these presentations are less likely to be prepared and delivered using visualization or business intelligence tools.
They are more likely to be recorded, archived, and shared than small team meetings, as the latter would be less coherent to an outside audience who does not share the team's context. 

The focus and frequency of communication between cross-functional groups in an organization impact the extent to which presenters reuse or appropriate existing presentation materials.
P13, a senior data strategist at a software company prepares reports, dashboards, and presentations for various client departments across the organization, and he delivers presentations twice a week.
Depending on the audience, he describes how the 
\revision{depth of analysis conveyed by visualization} can vary considerably:  \textit{``If you go to my engineering team, they don't really care about visualization as much as they care about the meaning that comes from that visualization; for the sales team, it is different; you are just looking at quotas: have you met the goal or not?''}

Another frequent presenter is P07, a marketing manager for a company that makes educational materials.
Like P13, she acts as a liaison between cross-functional groups, connecting the company's sales and marketing groups, presenting at biweekly sales meetings, large annual meetings, and occasionally to the owner. 
She cited the ability to present data in different ways as being critical to her presentation delivery, to engage her audiences with visualization design choices beyond conventional statistical charts, such as donut charts, waffle charts, and radial bar charts.
She explains, \textit{``I don't want to always show the same things; everybody knows how to make a bar graph; everybody knows how to make a pie chart; everybody knows how to make a line graph; sometimes I want to show them something different.''}

In contrast, a consistent audience and a less frequent presentation cadence sees P09 reusing more of his materials from one presentation to the next, \revision{and using a smaller set of chart types}.
As a marketing manager for a multinational coffee retailer, he delivers presentations on a quarterly basis about consumers' brand recognition across the competitive landscape
to an audience of regional executives and colleagues in the broader marketing wing of the organization.
Each quarter, he is able to retain the same presentation structure and incorporate the same charts that have become familiar to his audience. 
As a result, most of the presentation preparation work is spent updating the presentation materials with new customer data.
A slower cycle of refinement takes places over the course of two to three quarters, replacing charts as needed and identifying which essential values should be retained.


\subsection{Formal Presentations as \textsl{\revision{Recitals}}}
\label{sec:scenarios:formal}

Moving further along the dimensions of formality, preparation effort, and organizational distance between presenter and audience, this scenario is the one we most often associate with performative presentations of data.
These are clearly structured and rehearsed, uninterrupted during their delivery, and can require months of preparation.
As mentioned in \autoref{sec:rw}, our familiarity with this scenario is the result of our exposure to public-facing presentations about data at conferences like TED or in broadcast media.

Within organizations, such presentations can afford more interactivity with the audience relative to those occurring in the public sphere. 
The goals of these intra-organizational presentations also differ from the educational and inspirational imperatives of public presentations; for instance, we heard about presentations to celebrate milestones and announce new targets for a sales organization (P05 and P06), to provide evidence of an initiative's success and maintain trust (P01), or to persuade executives to green-light a new initiative (P09).

We previously encountered manufacturing association analyst P03 and marketing manager P09 in the context of smaller presentations;
we learned that they also periodically prepare presentation materials for their supervisors to present to executives and customers, which might include data tables, charts, and slides.
In both cases, the charts are typically provided as drafts and are reproduced either by an internal graphic design team or by an external agency.

One noteworthy collaborative work dynamic was described by P05 and P06, a data analyst and communications specialist who work together at a software company, where they prepare quarterly and annual presentation materials for the executive vice president of sales.
In any of these 90-minute presentations, it was typical for 20 minutes to be spent on \textit{`the numbers'}, which are initially visualized in Tableau workbooks. 
They described how they review these workbooks with \textit{``sales leaders and people from sales operations, and as soon as everyone's tied off on the validity and the accuracy of the numbers, we screenshot what we have here and put that image directly into a slide.''} 
Occasionally, they recreate some charts in PowerPoint, which they saw as offering superior formatting capabilities; they described 
\textit{``pushing the formatting down to PowerPoint so that we can make this slide as beautiful and as legible as possible.''}
Thankfully, the agenda of these presentations seldom changes from one quarter to the next, which spares P05 and P06 from the tedium of reformatting new slides and charts; instead, they can focus on updating and verifying the new quarter's data.
Also of note was that despite being the speaker, the vice president delegates the control of the slides to P05 during the delivery of the presentation.

Similarly, P09 presents to his senior management at least once a quarter, but also occasionally works with the CEO to prepare presentations to the owner family of the company. These presentations are highly produced and rehearsed, prepared over the course of several months, with visualization and graphic design being outsourced to an external firm. They usually concern changes in strategy or product lines and are delivered by the CEO, but P09 is on hand to answer questions.

At smaller organizations, it is more common that a single person will both prepare and deliver formal presentations.
For instance, P08, a product manager at an employee training service, collects and combines data from sales, customer satisfaction surveys, and service usage into presentations that serve to inform the company roadmap. 
She presents quarterly, to \textit{``our organization's leadership}: [the] \textit{president of the company, VP of strategy, VP of product, etc.''}

\section{Additional Characteristics of Presenting Data}
\label{sec:characteristics}

We expand on three characteristics of presenting data within organizations that span the three scenarios: how often presentations take place, the tools used, and the visualization design choices taken. 

\subsection{Presentation \revision{Frequency}}
\label{sec:characteristics:cadence}

Presentations in organizations are seldom one-off events.
All of our interviewees present at least quarterly, and they present at different cadences to different audiences, whether it be to bring updates to their teams or managers or to review performance relative to goals. \revision{This contrasts with an earlier finding by Kandel~\etal~\cite{kandel2012enterprise} that suggested work across teams to be an exception rather than the norm; our interviewee pool was not limited to analysts, which could explain this discrepancy.}
For instance, P13 presents twice weekly to different teams in his organization, while P04 at times meets with her team several times a day, saying \textit{``you don't want to see my calendar!''}

Few of these presentations are built from scratch, since the audience expects to see updates on the numbers and situations they saw in the previous presentation, so they usually require reviewing and updating existing slides and charts.
Even for an individual presentation, the data can change multiple times while it is being prepared. 
Quarterly sales results, as in the case of P05 and P06, take weeks to consolidate and are subject to change up until the presentation delivery. 
As a result, they occasionally need to update all of the charts in a presentation multiple times. 
While not every chart necessarily changes with every update, they still replace all of them so as not to miss any detail and risk presenting outdated data.

\begin{figure*}[hb!]
  \centering
  \vspace{-3mm}
  \includegraphics[width=\textwidth]{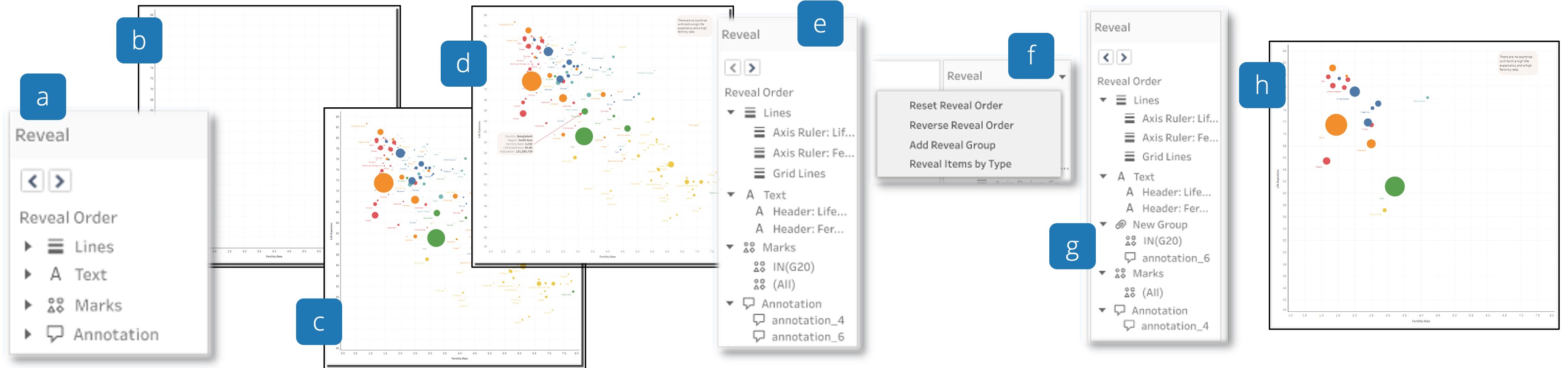}
  \caption{This example from the progressive reveal design probe demonstrates the reveal of a scatterplot. A reveal control interface (a) indicates an initial ordering of lines \& axes (b), marks (c), and annotations (d). This ordering can be expanded (e), appended to with a new reveal group (f), and reordered (g), resulting in a new sequence that interleaves marks and annotation (h) before arriving at the final reveal state (d).}
  \label{fig:progressive-reveal}
\end{figure*}

\subsection{Presentation Tools}
\label{sec:characteristics:tools}

The static nature of slides forces presenters to constantly switch between the presentation and data tools they need to prepare and deliver presentations.
P18 succinctly summarized this split, saying \textit{``I get this sense of failure every time I'm going into PowerPoint''}.

For instance, P10, who works for a healthcare consulting firm, described presentations about hospital utilization that drew upon models built in Excel.
While she favored PowerPoint for authoring sequences of animations
(\textit{``we love animations and just slowly telling a story little by little''}), she reported toggling from PowerPoint to Excel to respond to the audience's needs.
Similarly, P01, a productivity manager at a software company, reported presenting with PowerPoint decks containing screenshots from Tableau.
In lower-stakes presentations, he would pull up Tableau in response to questions or when an audience member \textit{``fundamentally challenged our base-level assertions.''} 
He would not switch between Tableau and PowerPoint in higher-stakes meetings with executives, explaining that \textit{``the biggest value} [\ldots] \textit{is time,''} and the cost of switching did not seem worth it.
He cited inertia for PowerPoint dependence, but also a need to collaborate with others using PowerPoint. 

While not surprising, it should be noted just how poorly suited existing slideware tools are for data: presenters can either create their charts in a visualization tool and export screenshots for their slides, or they can recreate their charts from scratch in the presentation tool, albeit without the connection to the original data sources, as slideware tools lack data shaping and processing capabilities.
Either process creates static snapshots that are frozen in time and will need to be updated when the data changes. 
Both processes are tedious, error-prone, and require repeated labor to check and update the data; mistakes could propagate throughout the presentation materials and lead to incorrect audience interpretations. 
P08 captures this tedium by describing how she has to \textit{``make a chart, copy that chart, put it in the slide deck, write up all of my information that I need,''} lamenting the inability to do \textit{``it all in one place.''}

\subsection{Visualization Design Choices for Presentations}
\label{sec:characteristics:techniques}

Presentation scenarios have different goals from analysis, which lead to different visualization design choices~\cite{kosara2016presentation}. 
As P09 put it, his job is to \textit{``synthesize that [data] and make it snackable.''}

\revision{Our interviewees use a variety of chart types in their presentations, from familiar bar and line charts to less familiar ones like donut and radial bar charts.
However, we identified no clear correspondence between chart type and presentation scenario.}
Several interviewees mentioned using specific chart types to attract the audience's attention and to present data in ways that they see as being more aesthetically pleasing. 
Typically, this meant charts with circular elements, such as donut charts, bubble charts, and lollipop charts; P01 uses the latter when he \textit{``wants to break up an otherwise boring series of charts''}.
Similarly, P09 said that \textit{``if you could present ordinary data in extraordinary ways, it would come across much better and you'd really drive home your point}.'' 
In his own presentations, when showing one bar chart after another, he remarked how \textit{``some people are half asleep}.'' 
He told us that he \textit{``hate}[s] \textit{bar charts''}, describing them as dated (\textit{``bar charts are very 2001''}), adding that \textit{``sometimes you need to do it in different ways to keep the attention.''}

We also asked about interviewees' solutions for constructing these presentation-oriented charts.
P07 uses Venngage~\cite{venngage}, which allows her to create simple infographic-like charts from templates that include donut charts, waffle charts, and radial bar charts. She uses these templates as inspiration for her to create a greater diversity of charts. 
Others outsource the construction of such charts; for instance, P09 works with an outside design firm on some of the presentations he prepares for his CEO to deliver, which often feature highly stylized 3D charts.

\section{Performative Presentation Design Probes}
\label{sec:probes}

To anchor conversations around presenting data to an audience, we developed three sets of prototypes to serve as design probes, each exploring a different aspect of presenting data. 
For each prototype, we prepared a short video demonstrating the functionality. 
We consolidated these videos into a single presentation accompanied by a scripted explanation; we provide an edited recording as supplemental material~\cite{supplemental}.
As interactive wireframes, we built these prototypes using Tableau's Extensions~\cite{tableauext} and JavaScript~\cite{tableaujs} APIs, InVision~\cite{invision}, and PowerPoint. 
Given the concurrent time span of the retrospective and design probe interviews, these design probes do not directly address the three scenarios described in \autoref{sec:scenarios}. 
However, we discuss how they may apply to the three scenarios in \autoref{sec:discussion}.
\revision{Our thematic analysis of responses to these design probes resulted in twelve themes, indicated throughout this section as \textbf{bold paragraph titles}.}

\subsection{The Progressive Reveal of Charts and Dashboards}
\label{sec:probes:reveal}

Inspired by the animation pane in PowerPoint and similar controls in other slideware tools, we considered analogous experiences for selectively revealing content within a chart or dashboard. 
We developed four prototypes, each based on a different reveal scenario.

We presented the progressive reveal prototypes in order of increasing complexity, beginning with the reveal of marks within a single chart: revealing marks according to a specified categorical dimension; controlling the rate at which line segments in a line chart are revealed by specifying breakpoints along a temporal dimension; and generalizing the reveal of marks by specifying breakpoints along a continuous dimension. 
Next, we demonstrated progressive reveal at the level of chart components, beginning with an initial reveal order of zero lines and grid lines followed by axis names and tick labels, which were in turn followed by marks and annotations. 
Finally, we demonstrated the grouping of chart components to change their reveal order, such as by interleaving the reveal of annotations with the reveal of marks, as illustrated by the sequence depicted in \autoref{fig:progressive-reveal}.

\bstart{Accessible presentations and attention management}
Interviewees saw progressive reveal controls as a way to ensure an accessible audience experience, particularly when presenting data to a global audience including speakers of different languages, as is common in multinational organizations. 
P05 suggested that a presenter \textit{``could use} [progressive reveal of annotations] \textit{or not use it depending on their audience,''} thereby making on-the-fly choices to selectively reveal annotations during a presentation instead of showing all of the annotations. 
Progressive reveal controls would encourage presenters to assume \textit{``a nice pace''} [P15] and to acknowledge that some audiences \textit{``need a little bit more help to stay engaged,} [\ldots] \textit{a little extra guidance''} [P05]. 
P16, a self-professed \textit{`customization freak'} who routinely presents complicated financial data, remarks that despite its complexity, this data tends to have a logical flow: \textit{``if I could reveal things in that meaningful flow so that people can see how things relate to each other, I think that could be really useful.''}
Finally, the inverse of progressive reveal was seen as an equally viable strategy for capturing attention. 
P14 suggested progressively removing or de-emphasizing content within a chart to iteratively focus attention: \textit{``you might subsequently take one thing away and then reveal another} [\ldots] \textit{or bring something to the foreground in a way that if you currently highlight a mark by selecting it, essentially pushing everything else to the background.'' }

\bstart{Flexibility and scope of reveals}
Another benefit of having control over the reveal of data is flexibility, of being able to tell a story differently to different audiences, or to selectively reveal specific data in response to audience questions: \textit{``there are times when you want to show one small piece of data} [\ldots] \textit{and other times you want to be able to see the full picture, being able to have more flexibility can be really helpful in the presentation''} [P05]. 
One suggestion pertaining to flexible reveal had to do with the granularity of comparison; specifically, P15 considers the use case of a faceted or \textit{small multiples} chart, where \textit{``you start off with one} [facet]\textit{, and then you bring in another.''}
While faceted charts are commonly used to support comparison judgments, P16 points out that it is also important to allow for comparisons at the level of marks, such as by revealing sets of marks within a chart independently: \textit{``grouping and ordering things could have a totally different impact on the way that the} [audience] \textit{would see that, especially when it comes to being able to group the marks.''}

\bstart{Reveal for suspense and drama}
Flexible controls for revealing data could allow presenters to anticipate desirable affective responses from their audience. 
P17 commented on the suspense inherent to presenting sales performance during the final week of a financial quarter: \textit{``It's always sort of a slow build} [\ldots] \textit{when we've had a big week, having that big reveal speaks to that.''} 
P17 prepares presentation materials for P18, a lively storyteller who immediately foresaw the application of progressive reveal to his quarterly presentations: \textit{``here's last year; here's this year; look at where we're going} [\ldots] \textit{we know we were struggling --- and then bang! --- Here comes the third quarter; look at what we did in the last week.''}  
He wanted \textit{``the ability to, in a simple way, create a level of drama and tell the story better; I would use that a lot.''}
As pacing is critical to drama~\cite{Chu2016}, P17 lamented the absence of an ability \textit{``to control it really slowly, then go fast, then slow it down at various points.''} 
With suspense at stake, reveal controls should prevent situations where \textit{``if your mouse slips, you have to do it again,''} and the dramatic potential is lost.

\begin{figure*}[b!]
  \centering
    \vspace{-3mm}
  \includegraphics[width=\textwidth]{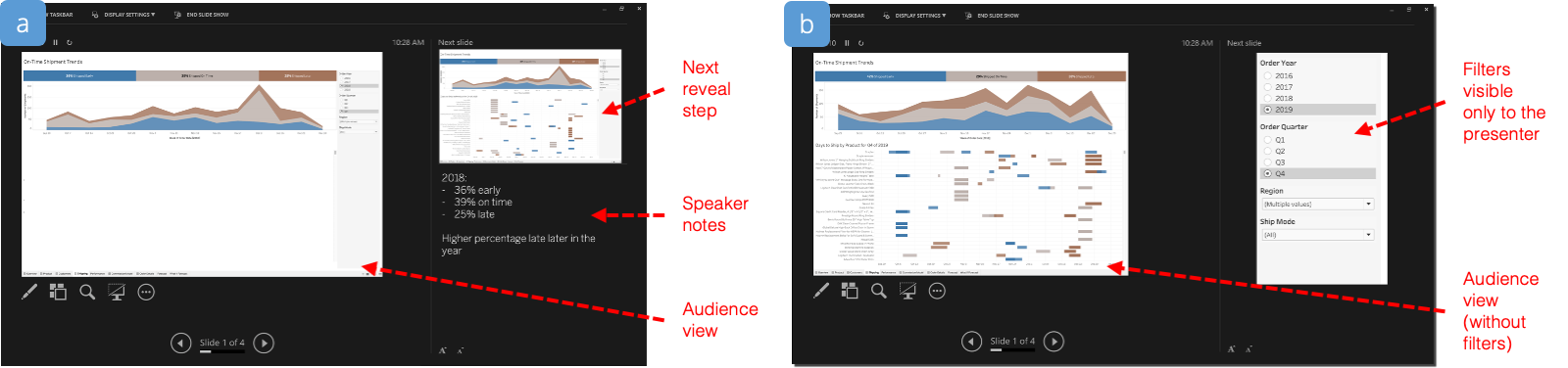}
  \caption{In one variant of the second-screen design probe (a), the presenter view shows the audience's view of a dashboard, the next reveal state of the dashboard, and speaker notes. In the second variant (b), the interactive filter controls originally shown as part of the dashboard are no longer shown to the audience; they are now only accessible to the presenter.}
  \label{fig:presener}
\end{figure*}

\subsection{Second-Screen Controls for Presenting Data}
\label{sec:probes:presenter}

Those who use slideware tools tend to be familiar with a presenter view, visible only to the presenter when extending their desktop. 
Our own second-screen presenter view design probe extends this idea to presenting data. 
We modified a screenshot of a PowerPoint presenter view to show the current and next reveal state of a chart or dashboard. 
In \autoref{fig:presener}a, the preview on the left is a dashboard as it is currently shown to the audience on the primary presentation display, while the smaller preview on the right displays the next reveal state; speaker notes also appear here. 

Beyond progressive reveal, presenters may need to interact with a chart or dashboard during the presentation, such as to respond to questions from their audience. 
\autoref{fig:presener}b depicts filter controls being removed from the dashboard so that they are only visible and accessible to the presenter. 
A filter panel now appears on the right of the presenter's display. 
This probe supposes that the audience does not need to see how a change in a dashboard is specified; they only need to attend to the result of the change.

\bstart{Presentations, not tool demonstrations}
Our interviewees who were accustomed to presenter views in slideware were enthusiastic about extending the concept to presenting data. 
Moving filter controls from the audience display to the presenter's display could serve as speaking cues in and of themselves according to P18 (\textit{``having the filters sitting in there; that almost gives you the prompts that you want''})
With this functionality, one can really change the feel of a presentation; by taking
\textit{``the filter off of the view and not having your mouse show}, [it] \textit{really feels like a presentation versus a demo''} [P05].

\bstart{Presenter tools beyond filters and notes}
P15 advocated for pointing and highlighting controls, referring to \textit{``the laser that you have in PowerPoint allows me to kind of go around the screen,} [\ldots] \textit{enlightening other areas} [\ldots] \textit{a small indicator to show where you're looking.''} 
P18 spoke of highlighting with shape annotations: \textit{``I will put up a} [chart] \textit{and then I'll throw up a red circle on the bit that you want to look at, to shine the spotlight on that.''}  
However, as P02 points out, \textit{``annotations obscure things,} [\ldots] \textit{you have to find a place for them.''} 
Ephemeral annotations could alleviate this concern, as suggested by P01: \textit{``something where I could hold control and drag a box or hold shift and make circles or make highlights or be able to annotate quickly, with some sort of hotkey to annotate a mark or something on the fly.''} 

\bstart{Presenter tool templates}
One question to arise was whether a presenter view could be templated and re-used across presentations: \textit{``I'd really like a standard or cookie-cutter reveal path; can I take that and move it to another} [presentation] \textit{easily?''} [P01]. 
Still thinking within a slideware mindset, others asked whether the presenter controls would be \textit{``bespoke for each slide''} [P14] or for each chart or dashboard: \textit{``I would want some form of consistency from slide to slide so that I knew what to expect and where to expect it''} [P15]. 

An alternative approach to presenter view templates would be to modify the audience view, not the presenter view.
This suggestion comes from P02, who routinely presents screenshots from Tableau in PowerPoint.
He questioned the need for a custom presenter view akin to what is currently provided by the latter: \textit{``the best thing would be just to have a regular Tableau view on your local monitor and then on your presentation monitor, it would be whatever simplified display you elect.''} 
By using a familiar data analysis interface as a presenter view, the controls would remain in familiar places while presenting. 

\bstart{Little concern over changes behind the curtain}
We speculated that hiding filter controls from an audience might reduce trust in the presenter, or the presenter might give the impression that they are hiding something from the audience. 
Overall, our interviewees did not share this concern. 
P01 expressed concern as someone who focuses on demo and training scenarios: \textit{``if someone doesn't have that context,} [\ldots] \textit{they can make bad decisions or make wrong inferences based on change blindness.''} 
Aside from these remarks and similar ones about demo scenarios made by P17, our interviewees were not concerned by hiding these interactions. 
It is expected that an effective presenter will effectively narrate and highlight changes as they occur; the audience need not see the interface where these changes are specified, unless it is a product demo. 
Moreover, visible filters and interface widgets might introduce distractions and derail the conversation: \textit{``it's a little clunky, as you might have quite a lot on there; you're losing real estate} [to] \textit{a bunch of filters that you may or may not use; you're losing the aesthetics of the thing''} [P18].

If it were possible to present using a visualization or data analysis tool instead of a slideware tool, the audience experience need not change to a great extent. 
P14 remarked: \textit{``I wouldn't show that all those controls down the side of my PowerPoint deck to demonstrate what I've filtered at any one time. I'm responsible for explaining if I've changed anything from one view to the next; I think the onus is on} [me] \textit{.''} 

In the event that audiences do experience change blindness~\cite{simons2005change}, or if a presenter fails to effectively highlight the changes or refer to them in their spoken narrative, P02 suggested showing ephemeral captions that indicate what change has just occurred in succinct and plain English, such as \textit{``Fiscal Quarter filter updated to Q4 (from Q3).''} 

\bstart{Tool-switching and material unrelated to data}
Second-screen presenter controls for visualization or data analysis tools would be particularly ideal if the entire presentation could be given using these tools, but as P05 points out, many presentations involving data also tend to incorporate static text and other media such as images and video.
This content is easy to present in slideware but it can be tedious to present in visualization or analysis tools.
She asked: \textit{``how do you do both without having to switch platforms?} [\ldots] \textit{That's the part that's hard, especially for virtual presentations} [\ldots] \textit{going in and out of different platforms when you're presenting (especially virtually) is really tricky; you take a risk of something going wrong in that transition point.''}

\begin{figure*}[b!]
  \centering
  \vspace{-3mm}
  \includegraphics[width=\textwidth]{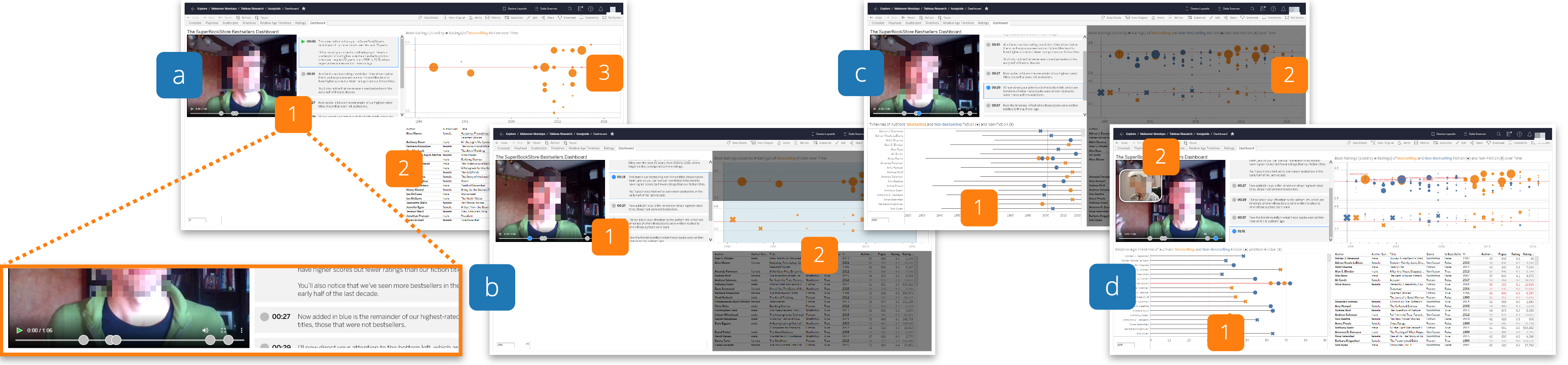}
  \caption{The third design probe illustrated the coordination of pre-recorded video of a presenter with an interactive dashboard. In (a), the video timeline (1) is augmented with a transcript and waypoints corresponding with states of the dashboard, which in this example initially contains a table (2) and a scatterplot (3). The first waypoint (b) advances the transcript (1) and spotlights part of the scatterplot (2). The second waypoint (c) adds a new timeline view to the dashboard (1) while dimming the other views (2). The final waypoint (d) changes the scale of the timeline view (1) and adds an image annotation (2). At any point, the viewer can pause the video, jump to a different waypoint, and interact with the dashboard.}
  \label{fig:tourguide}
\end{figure*}

\subsection{Coordinating Video with Interactive Visualization}
\label{sec:probes:tourguide}

Our final design probe addressed the gap between synchronous and asynchronous viewing of presentations about data. 
Many presentations and meetings in organizational settings are recorded, though the downside of disseminating a recording is that it does not allow viewers to interrupt the presenter and ask questions about the data. 

To bring aspects of synchronous presentations to asynchronous consumption, we prototyped interfaces where viewers can watch a prerecorded video that is coordinated with interactive visualization. 
Videos would have one or more associated waypoints: timestamped events along the video's timeline linked with different states of a chart or dashboard. 
\autoref{fig:tourguide} depicts one of these prototypes, where a video of a presenter is embedded within a visualization dashboard. 
In this example, an initial dashboard state (\autoref{fig:tourguide}b) contains a video with five waypoints (1: represented as grey circles along the video's timeline), a table (2), and a scatterplot (3). 
Text blocks adjacent to the video could either be a verbatim transcript of the presenter's monologue or a summary caption for each corresponding state of the dashboard. 
Different dashboard events occur at these waypoints, which include spotlighting parts of a dashboard (\autoref{fig:tourguide}b-2), changing filters (\autoref{fig:tourguide}b-2), adding (\autoref{fig:tourguide}c-1) and replacing (\autoref{fig:tourguide}d-1) dashboard content (\autoref{fig:tourguide}c-1), or adding annotations such as text boxes, reference lines, and images (\autoref{fig:tourguide}d-2). 
Unlike a recorded video, the audience can interact with the dashboard during playback, or they can navigate to states that are associated with a particular waypoint.

This presentation consumption experience could replicate another aspect of live events: the ability of a presenter to incorporate absolute deictic references in their monologue. 
Many presentation recordings do not incorporate a presenter's video, or if they do, the video is often relegated to a small thumbnail. This is unfortunate given the importance of visual and nonverbal cues that people rely upon for successful communication: facial expressions, hand gestures, and shared spatial references. 
In our prototype, the presenter video is in a prominent and fixed location within the dashboard. 
As a result, the presenter can use gestures and spatial references like \textit{`the chart to my left'} or \textit{ `the chart below my video'}. 
The aim of this design probe was to illustrate a restoration of some of the human connection between presenter and audience that people rely on in live in-person presentations, but that is largely lost in remote presentations.

Finally, coordination between video and interactive visualization may extend beyond dashboards. 
The final part of this design probe illustrated alternative coordination experiences embedded on the web or as a form of content consumable from a mobile business intelligence application (see supplemental video~\cite{supplemental} from 2m50s).


\bstart{A timely yet unexpected solution}
P18 stated that this design probe \textit{``feels very here and now, a world in lockdown} [\ldots] \textit{we're getting used to seeing each other on screen rather than in person.''} 
P15 echoed this: \textit{``you're adding the human element; it's something that is becoming more interesting now in relation to the pandemic} [\ldots] \textit{a lot of us are in our social isolation; video is something I've been thinking about.''} 

Despite encouraging comments about the timeliness of this design probe, most of our interviewees do not record presentations solely for asynchronous consumption. 
Instead, they disseminate blog posts, newsletters, emails, as well as recordings and slides after a live presentation. 
Despite being unexpected, this design probe did not strike our interviewees as being outlandish.
P18 had previously given thought to reworking his weekly status email as a video, while P14 remarked that \textit{``before} [today]\textit{, I certainly never considered this use case, but I can see huge value in this for my organization.''}

\bstart{Analytical onboarding vs.~linear presentation}
While four of our interviewees expressed enthusiasm about this design probe, others were more skeptical.
P13 said that he would \textit{``draw a distinction between a cool idea and a useful idea''}, explaining that \textit{``the cool idea is making a video of the presentation, [\ldots] but you can do that with a screen recorder. 
The useful idea would be to use this for teaching people how to interact with the dashboard.''} 

To this end, P01 and P11 could speak to disseminating recorded presentations solely for the purpose of teaching and onboarding. 
P11 produces video onboarding for his content published to Tableau Public~\cite{tableaupublic}: \textit{``I record audio or video of me talking} [about] \textit{how I interact with my dashboard and put it on YouTube, and then I put a link to that video in my Tableau Public dashboard, so that} [my audience] \textit{can watch it before they play with it.''} 
Meanwhile, P01 records onboarding videos for data analysis software, to be used in conjunction with an interactive web-based Learning Management System (LMS), which would include quizzes based on the contents of an attached dashboard. 
He remarked that this design probe brought all of these concepts together: a dashboard \textit{``functioning as a LMS – that's super exciting.''} 

Coordinating video with interactive visualization could replace existing onboarding processes that may not scale as organizations grow: \textit{``I have to reach out to} [colleagues] \textit{and schedule a one-on-one meeting and walk through things with them, do screen shares''} [P16]. 
As an alternative or complement to in-person onboarding, P14 mentioned that he \textit{``would be very keen to release a new dashboard and provide an instructional video as to how to engage with it.''} 

Beyond the applicability of this design probe for onboarding people to interactive dashboards, our interviewees did not take a stance on its applicability to linear narrative presentations. 
An exception was P11, who saw the interactive dashboard component as being inappropriate for linear storytelling: \textit{``by nature, storytelling, video, and animation is linear; they have a beginning and an end, regardless of whether the topic itself is linear or not; so that if you're using that linearity well, people shouldn't have to click too many things to understand your message.''} 
He also voiced concern over \textit{``the split attention effect''} that the video introduces, in which viewers are drawn to both the video of a dynamic presenter and to changes occurring in a chart or dashboard, which he likened to 
\textit{``a circus show where everything's flying all the time.''}
He suggested that voice-over narration would be sufficient, or at least an alternation between video playback and visualization content.

\bstart{A simpler loop-based viewing experience} 
Interviewees were more receptive to a pared-down browser-based version of this design probe relative to the original dashboard-based version; we show both in the supplemental video~\cite{supplemental}. 
With only one chart shown at any one time shown next to a presenter's video, P14 remarked that this simpler design was reminiscent of televised newscasts and could provide \textit{``a really good data journalism angle for} [his] \textit{service,''} upon which he imagined producing 3-minute videos that serve to explain data. 
Even P11, who was skeptical of the original design, claimed to \textit{``like this better} [\ldots] \textit{it's more straightforward; it gives less degrees of freedom to the audience} [\ldots] \textit{we can make sure that they stay on track''} [P11]. 
In his own presentations, P11 seldom shows more than two dynamic elements at a time, and if either element could be looped and / or slowed down, he would employ these strategies \textit{``to help deal with a cognitive overload.''} 
He likened this strategy to \textit{``how people reveal the secret behind a magic trick: they would play it from a different angle, and then play it again, but slower so that you can see the trick;} [\ldots] \textit{something similar could be done with data visualization, so that people can understand that better by being exposed to the same material over and over.''} 

\bstart{Authoring complexity}
Our interviewees also voiced concern over the perceived authoring complexity of coordinating video with interactive visualization. 
Among all of our interviewees, P01 was most accustomed to recording video content for onboarding people to data analysis software, using a suite of video recording and editing tools. 
He cautioned: \textit{``the barrier I see is if it's harder to use than something like Camtasia,}~\cite{camtasia} [\ldots] \textit{if I'm doing this regularly, it's always going to be short-term easier for me to use the things that I know rather than the new thing, so having that low barrier to entry or similar visual language''} would be necessary. 
Concern over learning a new presentation authoring tool was reiterated by P18, who stated that if it will take \textit{``three hours to build a ten minute presentation, I ain't gonna do it.''}

Though we explained that one would upload a prerecorded video to this design probe, P02 warned of a potential feature creep, in that those preparing presentations would expect full video editing capabilities in addition to visualization or dashboard authoring.
He was accustomed to using existing tools for coordinating a staged reveal of visual media for live presentations in legal settings, namely Sanction~\cite{santion} and TrialDirector~\cite{trialdirector}.
He suggested a simpler alternative to coordinating video with interactive visualization, 
a slide viewer component for a dashboard, where the current slide could set the state of the dashboard. 
Moreover, this coordination between slide and dashboard could be used both in asynchronous and synchronous presentations.

\section{Discussion \& Research Opportunities}
\label{sec:discussion}

\revision{We now step back to reflect on the limitations of our two studies, discuss how their findings complement one another, and point to opportunities for further research.}

\bstart{\revision{Limitations}}
\revision{Our findings should be viewed as a first step toward understanding the spectrum of preparing and delivering presentations with data. 
While our recruitment of professional interviewees was largely opportunistic, as opposed to being motivated by attaining an ideal saturation of results, we were nevertheless able to observe recurring themes across their responses. 
We acknowledge that there may be other scenarios that our community has yet to discover and characterize.}

\bstart{Connecting the design probes to the scenarios}
Controls for the progressive reveal of data (\autoref{sec:probes:reveal}) show promise for establishing a sense of drama and suspense, which suggests the more structured narrative presentations found in \revision{semi-improvisational performance and formal recital} scenarios.
For distraction- and interruption-prone presentation scenarios, presenters would likely benefit from the ability to control the pace of these reveals during presentation delivery.

Second-screen presenter controls (\autoref{sec:probes:presenter}) may be more generally applicable across scenarios. 
Even in more collaborative jam sessions, when these controls are manipulated solely by a moderator, others in the group would be less distracted by interface elements, particularly those who are less experienced with the tool. 

Finally, we learned that the coordination of video with interactive visualization (\autoref{sec:probes:tourguide}) may be inappropriate for presenting the linear narratives associated with \revision{semi-improvisational performance and formal recital} scenarios.
\revision{A simplified coordination between a video and a single chart may be useful in educational presentations, while a more richly coordinated experience may be an effective way of onboarding employees to analysis tools and processes, though these scenarios seem to be distinct from the three that we characterize in this paper.}


\bstart{\revision{From co-located to remote collaboration and presentation}}
\revision{
While we are excited by the potential of large displays~\cite{knudsen2019pade,langner2018multiple} or those affording touch- and pen-based interaction~\cite{lee2013sketchstory,zgraggen2014panoramicdata} to present and collaborate around data, none of our participants indicated prior use or access to these technologies.
Of more immediate concern to our interviewees was the need to communicate with remote audiences, a need reinforced by the onset of the pandemic.}
\revision{To this end, the visualization community could look to how some visualization practitioners~\cite{cwf2021,wu2021} have co-opted livestreaming platforms such as Twitch to demonstrate, teach, and experiment with visualization in response to a synchronous chat conversation with a live audience. 
Recent research also prompts speculation over what visualization-specific functionality might assist those presenting data on these platforms~\cite{chung2021,zhao2019data}.
In some remote jam sessions, functionality that provides collaborators a shared awareness of each others' interactions~\cite{neogy2020representing,schwab2020visconnect} could be beneficial, whereas this functionality may be less appropriate for more structured or formal presentations.}

\bstart{Better support for performative aspects of presentation}
Structuring presentations to make them more engaging and interesting \revision{is an art.} 
Some of our interviewees \revision{incorporate progressive reveals} quite effectively and expressed a desire for more control over them. 
\revision{This includes planning reveals during presentation preparation, controlling of the granularity of what is revealed (\eg~data points, chart elements, annotations) and when to reveal it, as well as how to control and execute new reveals during presentation delivery.}

Our interviewees also spoke of progressively highlighting specific data points. 
In live presentations, a presenter could point directly at an item or walk over to a screen and form a framing gesture with their hands. 
In remote presentations, these embodied deictic gestures and other expressive body language are not possible. 
We invite the research community to consider ways of replicating such gestures in remote settings and incorporating them into the execution of reveals.

\bstart{Integrated creativity support for presenting data}
Unlike in data analysis scenarios, the use of visualization in presentation contexts must engage the audience and capture their attention (see \autoref{sec:characteristics:techniques}). 
Our interviewees reported using chart types that many in our community would consider to be perceptually inefficient such as donut charts and lollipop charts, as well as templates with custom layouts and design elements, all as a means to avoid showing many similar-looking charts within a single presentation. 
This highlights a need for tools to recommend more varied representations beyond conventional charts, though these recommendations should still be grounded in the data at hand, so as to prevent the use of charts that are incompatible with the presenter's communicative intent.
Many slideware tools now provide creativity support in the form of recommendations for color palettes and slide layouts.
We call upon tool builders to unify this creativity support with recommendations for communicative visualization design choices within a single narrative presentation tool.

\section{Conclusion}
\label{sec:conclusion}

Seeing and understanding data is central to synchronous communication within organizations.
It should concern our community that visualization tools are seldom used for these activities to the extent that they could be. 
We corroborate prior findings that presenting data within organizations is typically a process of taking screenshots from visualization tools and dashboards and pasting them into slide decks~\cite{elias2012annotating,sarikaya2018we}.
Unsatisfied with this reality, we set out to understand why and how people communicate synchronously around data, using three design probes to elicit ideas that could lead us to realizing more appropriate ways of presenting and discussing data among colleagues.


Presentation is both a form of, and catalyst for, collaboration. 
Our research reveals a previously understudied scenario that blurs the distinction between presentation and synchronous collaboration: discussions among small groups of peers where a moderator initiates the meeting with prepared visualization content. 
The fluid transition between presentation, discussion, and collaboration in these jam session scenarios demands further research and is currently under-supported by tools, particularly given the tendencies of participants to interrupt, ask questions, make predictions, offer explanations, and come to decisions. 
The current separation between visualization tools for data analysis and slideware tools for presentation is counterproductive in this scenario.
\revision{Each focuses on one end of the process, with little concern for integrating both into a common workflow that does justice to both the data and the needs of those who prepare and deliver presentations.}


In closing, we challenge the visualization community to consider the presence of both a human orator and a live audience in communicative presentations of data, and to look beyond the formal \revision{recital} presentations of conference keynotes and TED talks.
We also encourage researchers to study individuals in organizations who do not self-identify as data analysts or data scientists, but who nevertheless interact regularly with data and visualization, particularly in activities related to communication and collaboration.

\acknowledgments{We thank Andy Cotgreave, Shay Koenig, Michelle Kosterich, and Melanie Tory for assisting with the interviews as well as Dan Cory, Britta Fiore-Gartland, and Maureen Stone for their feedback.}


\bibliographystyle{abbrv-doi-hyperref}

\bibliography{presentation}
\end{document}